\documentclass[reqno]{amsproc}

\usepackage{amsmath}
\usepackage[all]{xy}
\usepackage{amssymb,amsfonts}
\usepackage{amsthm}
\usepackage{amsmath}
\usepackage{amssymb}
\usepackage{pb-diagram}
\usepackage{amssymb, latexsym, amsmath, amscd}

\let\Bbb = \mathbb
\let\frak = \mathfrak

\theoremstyle{definition}

\theoremstyle{remark}

\numberwithin{equation}{section}
\newcommand{\comp}{\mbox{\small $\circ$}}

\begin{document}

\title[Quantization, Rotation Generators and Quantum Indistinguishability]
{Canonical Group Quantization, Rotation Generators and Quantum Indistinguishability}

\author{C.L. Benavides}
\address{Departamento de F\'isica, Universidad de los Andes,
Cra. 1E No. 18A-10. Edificio H. A.A. 4976,
 Bogot\'a D.C., Colombia}
\email{ca-benav@uniandes.edu.co}

\author{A.F. Reyes-Lega}
\address{Departamento de F\'isica, Universidad de los Andes,
Cra. 1E No. 18A-10. Edificio H. A.A. 4976,
 Bogot\'a D.C., Colombia}
\email{anreyes@uniandes.edu.co}
\thanks{Financial support from  Universidad de los Andes is gratefully acknowledged.}

\date{January 1, 1994 and, in revised form, June 22, 1994.}



\begin{abstract}
Using the method of canonical group quantization, we construct the angular momentum operators associated to
configuration spaces with the topology of ({\it i}) a sphere and ({\it ii}) a projective plane.
In the first case, the obtained angular momentum operators are the quantum version of Poincar\'e's vector, i.e.,
 the physically correct angular momentum operators for an electron coupled to the field of a magnetic monopole.
 In the second case, the obtained operators represent the angular momentum operators of a system of two
  indistinguishable spin zero quantum particles in three
spatial dimensions. We explicitly show how our formalism relates to the one developed by Berry and Robbins.
  The relevance of the proposed formalism  for an advance in our understanding of the
spin-statistics connection in non-relativistic quantum mechanics is discussed.
\end{abstract}

\maketitle

\section{Introduction}
The  connection between the spin of quantum particles and the statistics they obey is a remarkable example of a very
simply stated physical ``fact'' without the recognition of which many physical phenomena (ranging from the stability of
matter and the electronic configuration of atoms to Bose-Einstein condensation and superconductivity) would not have an
explanation. Nevertheless, the simplicity of the assertion ``integer spin particles obey Bose statistics and
half-integer spin particles obey Fermi statistics'' stands in bold contrast to its intricate physical origin. Indeed,
Pauli's proof of the \emph{Spin-Statistics Theorem} \cite{Pauli:40} (improving on earlier work by Fierz
\cite{Fierz:39}), showed that the spin-statistics connection was deeply rooted in relativistic quantum field theory.
The way to a rigorous proof of this theorem (from the mathematical point of view) was a long one and involved the
efforts of many people (see, for example,  the book by Duck and Sudarshan \cite{Duck1998}). The modern proof of the theorem,
in the framework of the general theory of quantum fields, is described in the book by Streater and Wightman
\cite{WightmanCPT:00}, where many references to original sources are given. There are also treatments of the theorem
within the algebraic approach to quantum field theory \cite{Haag1996}. An approach within Lagrangian field theory
(which is based on earlier work by Schwinger) and which makes use of  Lorentz invariance, but in a restricted sense,
has been pioneered by Sudarshan \cite{Sudarshan1975}, \cite{Duck:97}.
 Nowadays, the Spin-Statistics Theorem stands as a well-established result of theoretical physics.

In spite of all of these triumphs, many authors have been of the opinion that there might be alternative ways to prove
the Spin-Statistics Theorem, in a way that does not use the whole machinery of relativistic quantum field theory. This
``belief'' in a non-relativistic proof of the theorem has its origin (presumably) in the realization that the topology
of the underlying structures  of a quantum theory (e.g. symmetry groups, configuration spaces, gauge potentials, etc..)
may lead to the explanation and clarification of many features of the theory. For instance, from the work of Schulman
\cite{Schulman:68} it became clear that the path integral approach to quantization had to be modified if it was to be
applied to a multiply-connected configuration space. This led Laidlaw and DeWitt \cite{Laidlaw:71} to study the path
integral quantization of the configuration space of $N$ indistinguishable (spinless) particles in $\mathbb{R}^3$. They
arrived at the conclusion that there were exactly two inequivalent quantizations of such a system, one leading to Fermi
statistics, the other one leading to Bose statistics. The lesson was: If the indistinguishability of particles is taken
into account \emph{before} quantizing, then the Fermi-Bose alternative emerges (in three spatial dimensions) as a
consequence of the non-trivial topology of the configuration space. In this sense, one can dispense with the
symmetrization postulate, if quantum indistinguishability is taken into account right from the beginning. Parallel to
these developments was the work on quantization of non-linear field configurations by Finkelstein and Rubinstein
\cite{Finkelstein:68} where a general relation between kink exchange and rotations was established, using homotopy
arguments, that resembled the connection between spin and statistics (exchange of particles produces a phase
$(-1)^{2S}$ in the wave function, the same effect that a rotation through $2\pi$ has on the wave function of a single
particle of spin $S$). Leinaas and Myrheim \cite{Leinaas:77} reformulated the problem studied by Laidlaw and DeWitt in
a language very close to that of fiber bundles, obtaining the same results for three spatial dimensions (the Fermi-Bose
alternative). In addition, they found that in two spatial dimensions, the possible statistics were given  not by a
sign, but a phase factor (the so-called ``anyon'' statistics \cite{Wilczek1982}). Since then, a considerable amount of
work has been devoted to attempts at alternative, non-relativistic proofs of the Spin-Statistics Theorem.

Perhaps one of the most interesting and influential proposals that have been put forward in recent times is that of
Berry and Robbins \cite{Berry1997}. It is based on a generalization of Leinaas and Myrheim's work in which the spin
degrees of freedom are included in the treatment of two indistinguishable quantum particles. Although it does not lead
to a new proof of the theorem \cite{Berry:00}, it has inspired new  developments, both in mathematics and in physics.
For instance, the recent work of Atiyah and co-workers on configuration spaces \cite{Atiyah2001}, \cite{Atiyah2002},
\cite{Atiyah2002a} was motivated by the technical difficulties that appear when one tries to generalize the
Berry-Robbins construction to the $N$-particle case. Of more relevance for physics, their work seems to have given new
impetus to the non-relativistic spin-statistics issue (see, for example, \cite{Anastopoulos:02}, \cite{Peshkin2003},
\cite{Allen:03}, \cite{Peshkin2003a}, \cite{Sudarshan:03}, \cite{Kuckert:04}, \cite{Chruscinski:04}). It has also led
to several questions that, to the opinion of the authors, deserve attention. So, in order to advance in our
understanding of the problem, it is necessary to first settle those issues. One of them -a crucial aspect of the
Berry-Robbins approach- is the imposition of single-valuedness on the wave function. This condition has been studied in
detail by  the second-named author and collaborators \cite{Papadopoulos:04}, arriving at the conclusion that the
single-valuedness condition  is inconsistent with the assumption that the wave function be a section of a vector bundle
over the \emph{physical} configuration space. The global approach proposed in \cite{Papadopoulos:04} also allows to
explain why the proof presented in \cite{Peshkin2003}  fails \cite{Papadopoulos}.
 Another point, that will be the topic of this paper, has to do with the rotational properties of a quantum system of
indistinguishable particles. Recent work by Kuckert shows that it is possible to characterize the connection between
spin and statistics in terms of a unitary equivalence between the angular momentum operator of a single-particle system
and the angular momentum operator of a two-particle system, both operators being restricted to suitable domains
\cite{Kuckert:04}. We believe that a detailed analysis of that equivalence, which takes fully into account the topology
of the problem, could lead to interesting results. For this reason, in this paper we will construct the angular
momentum operators for a system of two indistinguishable particles of spin zero, using Isham's canonical group
quantization \cite{Isham:84}. We will see how, using Isham's method, we obtain structures ($SU(2)$ equivariance) that
are already present in the Berry-Robbins construction, though not explicitly. This is interesting, because one of the
advantages of the spin basis of Berry and Robbins (Schwinger construction) is that it allows for explicit computations.
Thus, we expect that the Berry-Robbins construction, suitably reinterpreted (as proposed in \cite{Papadopoulos:04}),
may in fact lead to an advance in our understanding of the spin-statistics connection.

 Let us
finish this introduction with a description of the contents of this paper. In section \ref{sec:isham} we briefly review
Isham's canonical group quantization method. In section \ref{sec:monopole} -as an example illustrating the dependence
of quantum observables on the topology of the configuration space- we then construct, using Isham's method, the angular
momentum operators for an electron coupled to the field of a magnetic monopole. In section \ref{sec:4} we
consider a system of two indistinguishable, spin zero particles. Again using Isham's method, we construct the
corresponding angular momentum operators. The paper finishes with some remarks and conclusions on section
\ref{sec:conclusions}.

\section{Canonical Group Quantization}
\label{sec:isham} Quantization of a classical system described by means of a symplectic manifold $(M,\omega)$ involves
the construction of a Hilbert space $\mathcal H$ and of a quantization map `` $\hat \,$ " allowing one to replace
classical observables $f$ (that is, smooth, real valued functions on $M$) by self-adjoint operators $\hat{f}$ acting on
$\mathcal H$. The quantization map is required to be real linear and injective, and should map constant functions to
multiples of the identity operator. Additionally, the Poisson bracket of two classical observables must be mapped to
the commutator of the corresponding quantum observables (Dirac's quantization conditions). It is well known that such a
full-quantization (which includes an additional irreducibility requirement) is, in general, not implementable (Van
Hove's theorem). Nevertheless,  there are several  quantization methods that allow one to pick a  subalgebra of the
Poisson algebra $(C^\infty(M),\{\, ,\,\})$ and to map it homomorphically to an algebra of operators, satisfying
physically and mathematically reasonable conditions. One of them, widely known,  is Geometric
Quantization\cite{woodhouse:80}. In this section we will briefly review a scheme developed by C. Isham \cite{Isham:84},
the method of Canonical Group Quantization. It has some similarities with
 Geometric Quantization and also uses some of the techniques developed by Mackey \cite{Mackey1968} and Kirillov \cite{Kirillov1976}.

The starting point of Isham's approach is the observation that, behind the usual quantum theory of a scalar particle on
$\mathbb{R}^n$, where the canonical commutation relations (CCR)
\begin{equation}
\label{eq:CCR} \left[\hat q^i, \hat p_j \right]= i\hbar \delta^i_j,  \;\;\left[\hat q^i, \hat q^j \right]=0=\left[\hat
p_i, \hat p_j \right]
\end{equation}
are satisfied, there is a group acting on the classical phase space of the theory by symplectic, transitive and
effective transformations. In fact, regarding $\mathbb{R}^n\times \mathbb{R}^n$ as an additive group, we see that the
action defined by
\begin{eqnarray}
\label{eq:RnxRn}
(\mathbb{R}^n\times\mathbb{R}^n)\times T^*\mathbb{R}^n & \longrightarrow & \;\;\;\;\;T^*\mathbb{R}^n\\
(\,(a,b),(q,p)\,)\;\;& \longmapsto &(q-a,p+b).\nonumber
\end{eqnarray}
has the properties mentioned above. That  (\ref{eq:CCR}) and (\ref{eq:RnxRn}) have something in common can be seen if
one considers the exponentiated (Weyl) form of the CCR. In fact, defining unitary operators $U(a)$ and $V(b)$ by
\begin{equation}
\label{eq:2.3} U(a):=e^{-ia\hat{p}}, \,\;\;V(b):=e^{-ib\hat{q}},
\end{equation}
one easily  checks that the position and momentum operators transform according to
\begin{eqnarray}
\label{eq:2.4}
U(a)\hat{q}U(a)^{-1} & = & \hat{q}-\hbar a,\\
V(b)\hat{p}V(b)^{-1} & = & \hat{p}+\hbar b.\nonumber
\end{eqnarray}
On a general configuration space $\mathcal{Q}$, there are no \emph{a priori} given position/momentum operators. For
example, the natural choice for the position operator on $\mathcal{Q}=S^1$ is the ``angle'' variable which, as is well
known, cannot be used as the basis for a quantum theory on $S^1$ \cite{Kastrup2006a}. In such cases, a good starting
point is the consideration of the symmetry groups of the classical configuration space. Once the appropriate group, the
\emph{canonical group} $\mathcal{C}$, has been identified, the construction of the corresponding quantum theory
proceeds by studying the unitary, irreducible representations of the group. One then sees that in the particular case
of $\mathcal{Q}=\mathbb{R}^n$ the CCR (\ref{eq:CCR}) arise as the unique (by virtue of the Stone-Von Neumann theorem)
solution of a purely geometric problem: the operators (\ref{eq:2.3}) provide an irreducible unitary representation of
the unique simply connected Lie group the Lie algebra of which is a central extension of the Lie algebra of the group
$\mathcal{G}=(\mathbb{R}^n\times\mathbb{R}^n, \,+\,)$. So, in this special case, the canonical group $\mathcal{C}$
turns out to be the Heisenberg group.

Keeping these preliminary remarks in mind, let us proceed to  describe the general scheme. It is based on a careful
analysis of the following  diagram:

\begin{equation}
\label{diag:1} \hspace{-2cm}\xymatrix{& &{0}\ar[r]&{\mathbb{R}}\ar[r]&{C^\infty(M,\mathbb{R})}\ar[r]^{\jmath}
&\mbox{HamVF}(M)\ar[r]&{0} .\\
& & & & & \mathcal{L}(\mathcal{G})\ar[u]_{\gamma}\ar@{-->}[ul]^{P}&}
\end{equation}
The meaning of the different terms appearing in (\ref{diag:1}) is the following.
\begin{itemize}
\item $M$ is a symplectic manifold. We are mainly interested in the case where it is a phase space, of the form
 $M=T^*\mathcal{Q}$, with  $\mathcal{Q}$ a
homogeneous space.
\item $\jmath$ is the map that assigns to each function $f$ on phase space (the negative of) its Hamiltonian vector field.
 Following the notation in \cite{Isham:84}, we shall write $\jmath(f)=-\xi_f$. The kernel of $\jmath$ is the set of
 constant functions on phase space, thus making the first row of the diagram a short exact sequence.
\item $\mathcal{G}$ is a Lie group, acting by symplectic transformations
on $M$. The Lie algebra of $\mathcal{G}$ will be denoted $\mathcal{L}(\mathcal{G})$.
\item The map $\gamma: \mathcal{L}(\mathcal{G}) \rightarrow \mbox{HamVF}(M)$ is the Lie algebra  homomorphism induced
by the $\mathcal{G}$-action.
\item Once the appropriate $\mathcal{G}$-action has been found (certain requirements must be met), one looks for a
linear map $P:\mathcal{L(G)}\rightarrow C^\infty(M,\mathbb{R})$ that is also a Lie algebra homomorphism.
\end{itemize}
The idea of the quantization scheme is the following. Let us assume that $P$ maps $\mathcal{L(G)}$ isomorphically onto
some Lie subalgebra of $(C^\infty(M,\mathbb{R}),\{\,,\,\})$.
 In this case  one can define  a quantization map by
 fixing a representation $U$ of the group and assigning to each function lying in
the image of $P$ the self-adjoint generator obtained from $U$  by means of $P^{-1}$. The existence  of a map $P$ with
the desired properties is not something obvious. There are obstructions coming from the fact that the map $P$
determines a class in the second cohomology group  of $\mathcal{L(G)}$ (with values in $\mathbb{R}$). Of course, there
might be many $\mathcal{G}$-actions on $M$ that could be considered. But the restriction will be imposed that the
diagram (\ref{diag:1}) must be commutative. The reason for the imposition of this restriction is that, given a (finite
dimensional) Lie subalgebra $\mathfrak{h}$ of $C^\infty(M,\mathbb{R})$,
 the Hamiltonian vector field
 that a function $f\in \mathfrak{h}$ generates, $\xi_f$,  gives place to a
one-parameter group, acting by symplectic transformations on $M$. If all these vector fields are complete, their
one-parameter groups will generate a group $\mathcal{G}$ of symplectic transformations and, if the mapping sending
$\mathfrak{h}$ into the set of Hamiltonian vector fields is injective, we obtain a Lie algebra isomorphism
$\mathfrak{h}\cong \mathcal{L(G)}$. On the other hand, given a symplectic action of a Lie group $\mathcal{G}$ on $M$,
there is a naturally induced map $\gamma: \mathcal{L(G)}\rightarrow \frak X(M) $. It is only if $\gamma(A)$ is a
Hamiltonian vector field that we can assign a function on phase space to the Lie algebra element $A$. For this reason,
the requirement that the image of $\gamma$ lies in $\mbox{HamVF}(M)$ must be imposed\footnote{This is automatically
satisfied if $H^1(M;\mathbb{R})=0$ or if $\mathcal{G}$ is semi-simple.}.
 The
idea is, therefore, to try to ``reverse'' this procedure: starting with a group $\mathcal{G}$ of symplectic
  transformations, we seek  a kind of ``inverse'' to the map $\jmath$. More precisely, we look for a Lie algebra
  homomorphism $P$ such that $\jmath\, \comp P = \gamma$.
In other words, $P$ must be a linear map satisfying
\begin{equation}
\label{eq:2.5} \left\{P(A), P(B)\right\} = P(\left[A,B\right])
\end{equation}
and
\begin{equation}
\label{eq:2.6} \gamma(A) = -\xi_{P(A)},
\end{equation}
for all $A$ and $B$ in $\mathcal{L(G)}$. Since every exact sequence of vector spaces splits, there is no difficulty in
finding a linear map $P$ such that the diagram commutes. The problem lies in (\ref{eq:2.5}). The condition
(\ref{eq:2.6}) fixes $P(A)$ only up to a constant (since $\ker \jmath=\mathbb{R}$) and in some cases it is possible to
adjust these constants so as to satisfy (\ref{eq:2.5}). But this is only possible if the cocycle defined by
\begin{equation}
z(A,B):=\left\{P(A), P(B)\right\} - P(\left[A,B\right])
\end{equation}
is also a coboundary. We thus see how the obstruction is measured by the second cohomology group of $\mathcal{L(G)}$.
In case the cocycle cannot be made to vanish by a redefinition of $P$, a central extension of $\mathcal{L(G)}$ by
$\mathbb{R}$ can be used to construct the desired map. As mentioned above, this is precisely the way in which the
Heisenberg group (and with it the CCR) arises from the action (\ref{eq:RnxRn}).

Once the appropriate canonical group $\mathcal{C}$ has been found\footnote{In some cases it is given by $\mathcal{G}$,
in others, it will be a Lie group whose Lie algebra is the above mentioned central extension of $\mathcal{L(G)}$.} a
quantization map can be defined by assigning to each element $P(A)\in \mbox{Im}P\subseteq C^\infty(M,\mathbb{R})$ the
self-adjoint generator corresponding to $A$ induced by a unitary, irreducible representation of the canonical group.
Since there may be inequivalent representations of the canonical group, we may also obtain different, inequivalent
quantizations of the same classical system. The general scheme can thus be divided in two main steps:
\begin{enumerate}
\item Find the canonical group $\mathcal{C}$.
\item Study the irreducible, unitary representations of the canonical group.
\end{enumerate}
In the particular case where $M=T^*\mathcal{Q}$, there is a natural place to start the search for the canonical group,
and it turns out that the representations can be constructed using Mackey's theory of induced representations. When $M$
is the cotangent bundle of some configuration space $\mathcal{Q}$, then every diffeomorphism on  it induces a
symplectic transformation, given by the pull-back operation on the bundle. Additionally, the exterior differential of
any smooth function on $T^*\mathcal{Q}$ induces a canonical transformation, by translations along the fibers. Since non
of these actions is transitive, it is necessary to consider both of them. The natural combination of these operations
can be regarded as coming from the  group action $\rho$ defined by  ($[h]\in$ $C^\infty(\mathcal{Q},\mathbb{R})/
\mathbb{R}$, $\phi\in \mbox{Diff}\mathcal{Q}$ and $l\in T^*_q\mathcal{Q}$):
\begin{eqnarray}
\label{eq:2.action} \rho_{([h],\phi)}(l):= \phi^{-1*}(l)-(d h)_{\phi(q)},
\end{eqnarray}
 provided the set $C^\infty(\mathcal{Q},\mathbb{R})/\mathbb{R}\times \mbox{Diff}\mathcal{Q}$ is endowed with the
structure of a semi-direct product. That is, $C^\infty(\mathcal{Q},\mathbb{R})/\mathbb{R} \rtimes
\mbox{Diff}\mathcal{Q}$ is the group with elements of the form $([h],\phi)\in
C^\infty(\mathcal{Q},\mathbb{R})/\mathbb{R}\times \mbox{Diff}\mathcal{Q}$ and with product
\begin{equation}
([h_2],\phi_2)\cdot([h_1],\phi_1)=([h_2]+[h_1\comp\phi_2^{-1}],\phi_2\comp\phi_1 ).
\end{equation}
Thus,  for $M=T^*\mathcal{Q}$, step (1) reduces to the problem of finding a suitable finite dimensional subspace $W$ of
$C^\infty(\mathcal{Q},\mathbb{R})/\mathbb{R}$ and a suitable finite dimensional subgroup $G$ of
$\mbox{Diff}\mathcal{Q}$. The group $\mathcal{G}$ of diagram (\ref{diag:1}) will then be given by $W\rtimes G$. At this
point, we refer the reader to Isham's article \cite{Isham:84}, for a thorough discussion of the method. For the
applications that will be presented in the next two sections, it will be enough to briefly comment on how the vector
space $W$ and the group $G$ make their appearance in the still more special case in which $\mathcal{Q}$ is a
homogeneous space. A brief discussion of the way in which the representations are constructed in this case will also be
presented at the end of this section.

Assume that  $R:G\rightarrow\mbox{GL}(W)$ is a representation of a  Lie group on a real, finite dimensional vector
space $W$. Then, a contragredient representation $R^*$ is naturally induced on $W^*$, by duality. It is defined as
follows ($g\in G$, $u\in W$ and $\varphi\in W^*$):
\begin{equation}
\left(R^*(g)\varphi\right)(u):=\varphi\left(R(g)u\right).
\end{equation}
Regarding $W$ as a configuration space, we have $T^*W\cong W\times W^*$. Using the representation $R^*$, one can
construct the semi-direct product  $W^*\rtimes G$ in the usual way. It is then possible to define a left action of
$\mathcal{G}:=W^*\rtimes G$ on $T^*W$, by setting
\begin{equation}
\label{eq:2.7}
 l_{(\varphi',g)}(u,\varphi):= \left(R(g) u, R^*(g^{-1})\varphi -  \varphi'\right).
\end{equation}
An element $\varphi$ of the dual space $W^*$ can be naturally regarded as a function $f^{\varphi}\in C^\infty(W,
\mathbb{R})$ by setting  $f^{\varphi}(u):= \varphi (u)$. The map $P$ is then naturally given by ($\tilde
A\equiv(\varphi,A)$):
\begin{eqnarray}
\label{eq:2.8}
P: \mathcal{L}(W^*\rtimes G)& \longrightarrow & C^\infty(T^*W,\mathbb{R})\nonumber\\
 \tilde A &\longmapsto & P(\tilde A):(u,\psi)\mapsto \psi\left(R(A)u\right)+\varphi(u).
\end{eqnarray}
As explained in detail in \cite{Isham:84}, all properties  that the diagram (\ref{diag:1}) must satisfy are fulfilled
in this case, with the exception that the $\mathcal{G}$-action is not transitive. This problem can be solved by
restricting the action to a $G$-orbit  of $W$, say $\mathcal{O}_v$, for some $v\in W$. This leads us directly to
configuration spaces of the form $\mathcal{Q}=G/H$ (if $\mathcal{Q}=\mathcal{O}_v$, then $H$ is the little group of
$v$). The action (\ref{eq:2.7}), as well as the map (\ref{eq:2.8}) can then  be restricted to
$G/H\cong\mathcal{O}_v\subseteq W$ and one can show that (\ref{eq:2.7}) is exactly of the form (\ref{eq:2.action}).

Thus, starting with a homogeneous space of the form $G/H$, one has to find a vector space $W$ on which $G$ acts, and
such that $G/H$ is a $G$-orbit. In  this case, the canonical group  can be chosen as $\mathcal{C}\equiv \mathcal{G}:=
W^*\rtimes G$. The unitary, irreducible representations of this group can be constructed using Mackey's theory of
induced representations. Generally, the resulting representation space will be the  space of square-integrable  sections
of a vector bundle $E$ over $\mathcal{Q'}=G/H$, constructed as an associated bundle to the principal bundle
$G\rightarrow G/H$, by means of an irreducible unitary representation of $H$. Here, the subgroup
$H$ is regarded as the isotropy group of a previously chosen element in the character group of $W^*$,
 $\mbox{Char}(W^*)$\footnote{Hence, $\mathcal{Q'}$ is a
$G$-orbit in $\mbox{Char}(W^*)$. In the examples we are interested in, these orbits coincide with the $G$-orbits
in $W^*$ and we can identify them, i.e. $\mathcal{Q'}\cong \mathcal{Q}$.}.
Integration of sections is carried out using the hermitian structure of the vector bundle and a $G$-quasi-invariant
measure $\mu$ on configuration space. The operators giving the representation of the subgroup $G$ of $\mathcal{C}$ are
constructed using a lift $l^\uparrow$ of the $G$-action $l$ on $\mathcal{Q}'$ to the corresponding vector bundle. This
lift is naturally induced by the right action of $G$ on the principal bundle. We are thus naturally led to consider
$G$-vector bundles over $\mathcal{Q'}$. Recall that a $G$-vector bundle is a vector bundle (with total space $E$) over a
$G$-space $\mathcal{Q}$, together with a lift $l^\uparrow$, i.e., a $G$-action on $E$ which is linear on the fibers and
such that the following diagram commutes ($g\in G$):
\begin{equation}
\begin{CD}
E @>{l^{\uparrow}_g}>> E \\
@V{\pi}VV @V{\pi}VV \\
\mathcal{Q'} @>{l_g}>> \mathcal{Q'}.
\end{CD}
\end{equation}
 If $\Psi$ is a section of the bundle (i.e. a \emph{wave function}), then the unitary operator $U(g)$ acts on it as
follows:
\begin{equation}
\label{repre:sec} (U(g)\Psi)(x) :=\sqrt{\frac{d\mu_g}{d_\mu}(x)}\; l^{\uparrow}_g \Psi(g^{-1}\cdot x),
\end{equation}
where $d\mu_g/d\mu$ is the Radon-Nikodym derivative of $\mu_g$ with respect to $\mu$. This $G$-representation can
be extended to the whole group $\mathcal{C}$ as follows (recall that $x$ is an element in a $G$-orbit of
$\mbox{Char}(W^*)$):
\begin{equation}
(V(\varphi)\Psi)(x):=x(\varphi) \Psi(x).
\end{equation}
 The infinitesimal version of these
relations  gives place to the corresponding self-adjoint generators, of which the angular momentum operators of a
particle whose configuration space is a sphere are one example, to which we now turn our attention.
\section{Magnetic monopole}
\label{sec:monopole}
\subsection{The classical problem}

In this section, we explore a simple but fundamental example: the problem of a point electric charge coupled to the the
(external) magnetic field of a fixed magnetic monopole. As is well known, the importance of this problem lies in the
fact that, in order for the quantum problem to be consistent, the electric charge of the particle must be quantized
\cite{Dirac:31}.

Classically, the dynamics of a particle of mass $m$ and charge $e$ coupled to the field produced by a magnetic monopole
of strength $g$ can be described by the Lagrangian
\begin{equation}
\label{eq:3.1} L(q, \dot q) = \frac{1}{2}m\dot{q}^2 + \frac{e}{c}\dot{q}\cdot A(q) ,
\end{equation}
where $q = (q_1,q_2,q_3)$ denotes the position of the particle. The vector potential $A$ must be chosen in such a way
that its curl gives a radial field. If $g$ denotes the magnetic ``charge'', then (using the notations $r=\|q\|$ and
$\hat r =q/r$) we require:
\begin{equation}
B:= \nabla \times A \stackrel{!}{=}g\frac{\hat r}{r^2}.
\end{equation}
This condition cannot be satisfied using a global gauge potential $A$. Thus, the Lagrangian (\ref{eq:3.1}) is only
locally defined. It is possible to give a global description of this problem, in the Lagrangian setting, but the
introduction of additional structures is necessary \cite{Zaccaria1983}. For our purposes, the local description will be
sufficient. We therefore introduce the following local expressions for the gauge potential:
\begin{eqnarray}
A^N(q)& := & \frac{g}{r (r + q_3)} (-q_2, q_1, 0), \nonumber\\
A^S(q) & := & \frac{g}{r (r - q_3)} (q_2, -q_1, 0) .
\end{eqnarray}
Using the general form of Noether's theorem\footnote{The general form of the theorem guarantees the existence of a
conserved quantity whenever the Lagrangian is invariant under a one-parameter group of transformations \emph{up to a
gauge transformation}.}, one can show that there are three conserved quantities, related to the action of the rotation
group on the configuration space. Since the Lagrangian is only locally defined, one has to compute the conserved
quantities using the two expressions for the gauge potential, $A^N$ and $A^S$. The conserved quantities obtained using
$A^N$ and $A^S$ turn out to be the same, up to an irrelevant constant term. They can be combined into a single vector
\begin{equation}
J = m q \times \dot{q} - \frac{eg}{c} \hat{r},
\end{equation}
that is to be interpreted as the \emph{angular momentum vector} of the particle.  In fact, working in the Hamiltonian
formalism (still in local coordinates) one obtains the following expression for $J$:
\begin{equation}
\label{eq:3.2} J = q \times (p - \frac{e}{c}A^N) - \frac{eg}{c} \hat{r} = L - \frac{eg}{c} K^N
\end{equation}
where
\begin{eqnarray}
\label{eq:3.8}
L &:=& q \times p, \nonumber \\
K^N &:=& \frac{1}{g}(q \times A^N) + \hat{r} = \frac{q - r \hat{z}}{r - z}.
\end{eqnarray}
Equation (\ref{eq:3.2}) can be used in order to compute the Poisson brackets of the components of $J$. The result is
$\{J_i,J_j\}=-\varepsilon_{ijk}J_k$.  Thus, the components of  $J$ satisfy angular momentum commutation relations and
are to be regarded as giving the correct expression for the angular momentum of the particle:
\begin{equation}
\label{eq:3.3}
J = L - \frac{eg}{c} K.
\end{equation}
In spite of the fact that the description of this system can only be given in local terms, the angular momentum is a
well-defined, global function. But, as is well-known, the situation changes drastically when we consider the quantum
version of the problem. There are different ways to analyze it, all yielding the same result: the wave function for an
electron coupled to the field of a magnetic monopole is a section of a line bundle over the configuration space. The
topology of this bundle is characterized by an integer number $n$ that relates magnetic and electric charge, giving
place to Dirac's famous result:
\begin{equation}
\label{eq:3.4} eg/c=\frac{n}{2}\hbar.
\end{equation}
Since the wave function is a section in some bundle, the corresponding angular momentum operators must be maps from the
space of sections to itself. A physically motivated and detailed analysis of this problem, involving the construction of
the angular momentum operators, can be found in \cite{Biedenharn1981}. There, the form of the angular momentum
operators is guessed from the classical expression, leading to an operator of the form $L-\mu K$, where $\mu= eg/\hbar
c$. The quantization condition (\ref{eq:3.4}) arises from a consistency requirement on the theory\footnote{This comes
from the fact that the wave functions, as well as the angular momentum operators, are defined only locally. The
consistency requirement imposed is that expectation values of the quantum operators, computed using the different local
expressions, must coincide in the overlap regions.}. In the next section we will arrive at the same result by applying
the canonical group quantization method to the magnetic monopole problem.

\subsection{The quantum problem}
The configuration space for the monopole problem is given by  $\Bbb{R}^3 \setminus\left\{0\right\}$. Since the monopole
field is spherically symmetric and we are only interested in the rotational properties of the system, we can  regard
the sphere $S^2$ as the configuration space on which the magnetic monopole problem is defined. Moreover, since the
sphere is a deformation retract of $\Bbb{R}^3 \setminus\left\{0\right\}$, the topological effects produced by both
spaces in the quantum theory are the same. In order to quantize, we want to think of the configuration space as a
homogeneous space. We choose the description of the sphere as the quotient $SU(2)/U(1)$. In this case, the canonical
group is given by $\mathcal{C}=(\mathbb{R}^3)^*\rtimes SU(2)$. Since we are only interested in obtaining the angular
 momentum operators, we only need to construct the $U$ operators, as defined in (\ref{repre:sec}). The Jacobian factor
 $d\mu_g/d\mu$ is equal to one in this case, because the measure is $SU(2)$-invariant. Hence, all we have to do
 is to choose an irreducible unitary representation of $U(1)$ in order to construct a vector bundle
 associated to the principal bundle $SU(2)\rightarrow SU(2)/U(1)$. The lift $l^\uparrow$ is naturally induced by the group
 product in $SU(2)$, as explained below.

Let
\begin{eqnarray}
\mathcal{U}_n: U(1) & \longrightarrow & \mbox{Gl}(\mathbb{C})\nonumber\\
          e^{i\phi} & \longmapsto     & \mathcal{U}_n(e^{i\phi}):= e^{i n\phi}
\end{eqnarray}
denote one of the unitary representations of $U(1)$ on $\mathbb{C}$, labeled by an integer $n$. The elements of the associated
bundle $\mathcal{L}_n:=SU(2)\times_{\mathcal{U}_n}\mathbb{C}$ are equivalence classes of the form $\left[(p,v)\right]$, with $p\in SU(2)$
and $v\in \mathbb{C}$. The equivalence relation is
\begin{equation}
(p,v)\sim (p\cdot \lambda, \mathcal{U}_n(\lambda^{-1}) v).
\end{equation}
Here,  $U(1)$ is regarded as the subgroup of $SU(2)$ consisting of all diagonal matrices of the form
$\mbox{diag}(\lambda,\bar\lambda)$, with $\|\lambda\|=1$. If we adopt the convention of denoting
the elements of   $SU(2)$ by tuples $(z_0,z_1)$ that represent matrices of the form
\begin{equation}
\left(
\begin{array}{cc}
z_0 & \bar z_1\\
-z_1 & \bar z_0
\end{array}
\right),
\end{equation}
then the right action of $U(1)$ on $SU(2)$, that is given by
\begin{equation}
\left(
\begin{array}{cc}
z_0 & \bar z_1\\
-z_1 & \bar z_0
\end{array}
\right)
\longmapsto \left(
\begin{array}{cc}
z_0 & \bar z_1\\
-z_1 & \bar z_0
\end{array}
\right)
\left(
\begin{array}{cc}
\lambda & 0\\
0 & \bar \lambda
\end{array}
\right)=\left(
\begin{array}{cc}
\lambda z_0 & (\overline{\lambda z_1})\\
-(\lambda z_1) & (\overline{\lambda z_0})
\end{array}
\right),
\end{equation}
can be equivalently expressed as
\begin{equation}
(z_0,z_1)\longmapsto (z_0,z_1)\cdot \lambda= (\lambda z_0,\lambda z_1).
\end{equation}
We will use these conventions in order to identify the bundle $SU(2)\rightarrow SU(2)/U(1)$ with the Hopf
fibration $S^3\rightarrow S^2$, when appropriate.  If in addition we consider the equivalence of
 $S^2$ with $\mathbb{C}P^1$, we can regard
the projection $\pi: SU(2)\rightarrow SU(2)/U(1)$ as the map $\pi((z_0,z_1))=\left[z_0:z_1\right]$.
Thus, the left action of $SU(2)$ on $S^2\cong \mathbb{C}P^1$ takes the following form ($g=(\alpha,\beta)$):
\begin{eqnarray}
\label{eq:3.5}
l : SU(2)\times \mathbb{C}P^1 &\longrightarrow& \mathbb{C}P^1\nonumber\\
\left(g,\left[z_0:z_1 \right]\right)&\longmapsto& l_g(\left[z_0:z_1 \right])=\left[\alpha z_0-\bar \beta z_1:
\beta z_0 + \bar \alpha z_1\right].
\end{eqnarray}
The left action of $SU(2)$ on itself given by the group product allows one to lift the action $l$ to the bundle
$\mathcal{L}_n$. It is given by the following expression ($g,p\in SU(2)$, $v\in \mathbb{C}$):
\begin{equation}
\label{eq:3.6}
l_g^{\uparrow}\left(\left[(p,v)\right]\right):=\left[(gp, v)\right].
\end{equation}
The action of the angular momentum operators on wave functions can then be obtained from the infinitesimal version
of (\ref{repre:sec}). Given that these operators act on the space of global sections of the bundle,
it is necessary, in order to be able to compare with the expressions known from the physics literature, to
obtain local expressions. Therefore, we will construct local trivializations for the bundle $\mathcal{L}_n$ and will then
compute the action of the infinitesimal generators, using local sections.
\subsection{Local description of $\mathcal{L}_n$}
The total space of the line bundle $\mathcal{L}_n = SU(2)\times_{\mathcal{U}_n}\mathbb{C}$ consists of equivalence
classes of the form $\left[\left((z_0,z_1),v\right)\right]$, with $(z_0,z_1)\in SU(2)$ and $v\in\mathbb{C}$.
The projection is the map $\pi_n:\mathcal{L}_n \rightarrow S^2\cong\mathbb{C}P^1$ given
 by  $\pi_n(\left[\left((z_0,z_1),v\right)\right]):=[z_0:z_1]$. In order to construct
local trivializations for this bundle, we start by  defining local charts, as follows.

Set
\begin{eqnarray}
U_N &=& S^2 \setminus \left\{N\right\} \, \, \, \, \, \, \, \, \, \,  \mbox{(sphere with north pole removed)}, \nonumber \\
U_S &=& S^2 \setminus \left\{S\right\}\,\, \, \, \, \, \, \, \, \, \, \mbox{(sphere with south pole removed)}. \nonumber
\end{eqnarray}
We define local charts using stereographic projections onto the complex plane. Let us denote
the local coordinates as follows:
\begin{eqnarray}
z: U_N & \longrightarrow &  \mathbb{C}\nonumber\\
  x & \longmapsto & z(x)
\end{eqnarray}
(stereographic projection from the north pole) and
\begin{eqnarray}
\zeta: U_S & \longrightarrow &  \mathbb{C}\nonumber\\
  x & \longmapsto & \zeta(x)
\end{eqnarray}
(stereographic projection from the south pole). Notice that if on $\mathbb{C}P^1$ we set
$U_0:=\{\left[z_0:z_1\right]\,|\, z_1 \neq 0\}$
 and $U_1:=\{\left[z_0:z_1\right]\,|\, z_0 \neq 0\}$, then we can define local charts
 that coincide with $z$ and $\zeta$ through the equivalence $S^2\cong \mathbb{C}P^1$, as follows:
\begin{eqnarray}
U_0 & \longrightarrow &  \mathbb{C}\nonumber\\
  \left[z_0,z_1\right] & \longmapsto & z:=\frac{z_0}{z_1}
\end{eqnarray}
and
\begin{eqnarray}
U_1 & \longrightarrow &  \mathbb{C}\nonumber\\
  \left[z_0,z_1\right] & \longmapsto & \zeta:=\frac{z_1}{z_0}.
\end{eqnarray}
Hence, $U_0$ can be identified with $U_N$ and $U_1$ with $U_S$. It will be convenient to keep in mind that
if $x$ is a point in the sphere with polar coordinates $(\theta,\varphi)$, then
\begin{equation}
z(x)= \frac{e^{i\varphi}\sin\theta}{1-\cos \theta}\;\;\;\mbox{and}\;\;\;\zeta(x)=
\frac{e^{-i\varphi}\sin\theta}{1+\cos \theta}.
\end{equation}
Local trivializations for the  bundle $\mathcal{L}_n$ can be defined in the following way.

Using the notation $g\equiv(z_0,z_1)\in SU(2)$, set

\begin{eqnarray}
\label{eq:3.7}
\varphi_N : \pi^{-1}_n(U_N) &\longrightarrow& U_N \times \Bbb{C} \nonumber \\
\left[( g, v)\right] &\longmapsto& \left([z_0: z_1], \left(\frac{z_1}{|z_1|}\right)^n v\right)
\end{eqnarray}
and
\begin{eqnarray}
\varphi_S : \pi^{-1}_n(U_S) &\rightarrow& U_S \times \Bbb{C} \nonumber \\
\left[(g, v)\right] &\longmapsto& \left([z_0: z_1], \left(\frac{z_0}{|z_0|}\right)^n v\right).
\end{eqnarray}
As can be easily checked, these maps are well defined, and provide local homeomorphisms.
From these local trivializations we obtain, for the   transition function $g_{SN}$,
\begin{equation}
(\varphi_S\circ\varphi^{-1}_N)(\left[z_0 : z_1 \right], w) =
\left(\left[z_0 : z_1 \right], \left(\frac{z}{|z|}\right)^n w\right),
\end{equation}
that is, $g_{SN} ([z_0 : z_1]) = (z/|z|)^n$. From this we see that the first Chern number of $\mathcal{L}_n$
is $n$. This means that an integer number, that at first was chosen to (partially) label a representation
of the canonical  group, also determines the topology of the bundle where the space of physical states is defined.

\subsection{Construction of the angular momentum operators}
Recall that the lifting $l^\uparrow$ of the $SU(2)$ action on the sphere to $\mathcal{L}_n$
 is induced by the corresponding lifting on the principal bundle. Therefore, $\mathcal{L}_n$ has the structure
 of a homogeneous  $SU(2)$-bundle:
\begin{equation}
\label{diag:2}
\begin{CD}
SU(2) \times_{\mathcal{U}_n} \Bbb{C} @>{l^{\uparrow}_g}>> SU(2) \times_{\mathcal{U}_n} \Bbb{C} \\
@V{\pi_n}VV @V{\pi_n}VV \\
S^2 @>{l_g}>> S^2
\end{CD} ,
\end{equation}
where $l_g{[g^{'}]} := [g g^{'}]$ and $l^{\uparrow}_g([g^{'},v])= [gg^{'},v]$.

What we want to do now is to use the local trivializations $\varphi_N$ and $\varphi_S$ to obtain
a local version of (\ref{diag:2}).
Using the map $\varphi_N$, we can obtain a local expression for $l_g^\uparrow$. The corresponding
map will be denoted $\sigma_g$ (see the diagram below):
\begin{equation}
\xymatrix @R=30pt @C=10pt{
   & & {SU(2) \times_{\mathcal{U}_n} \Bbb{C}}\ar[rr]^{l^{\uparrow}_g}
 \ar[d]^{\pi_n}\ar @/_/[dldl]_{\varphi_N}
 &
  & SU(2) \times_{\mathcal{U}_n} \Bbb{C}  \ar[d]_{\pi_n} \ar @/^/[drdr]^{\varphi_N}& \\
    & &
U_N \subset S^2 \ar[rr]^{l_g}& & U_N \subset S^2  & &\\
    U_N \times \Bbb{C} \ar[rrrrrr]^{\sigma_g} &   & & & & & U_N \times \Bbb{C}.  &
          }
\end{equation}
According to the diagram, we have $\sigma_g = \varphi_N \circ l^{\uparrow}_g \circ \varphi^{-1}_N$. Off course,
this map is well
defined only for elements $g\in SU(2)$ such that $l_g(U_N)\subset U_N$. Since we are interested in the
 infinitesimal generators of the group action, we will only consider group elements near the identity, so that this
 condition will always be satisfied.\\
Thus, for  $g=(\alpha,\beta)\in SU(2)$ and $([z_0:z_1], w)\in U_N\times \mathbb{C}$ we obtain:
\begin{eqnarray}
\sigma_g\left(([z_0:z_1], w)\right) &=&
(\varphi_N \circ l^{\uparrow}_g \circ \varphi^{-1}_N)([z_0:z_1], w) \nonumber \\
&\stackrel{(\ref{eq:3.7})}{=}&
(\varphi_N \circ l^{\uparrow}_g) \left(\left[\left((z_0, z_1), \left(z_1/|z_1|\right)^{-n} w\right)\right]\right) \nonumber \\
&\stackrel{(\ref{eq:3.6})}{=}&
\varphi_N \left( \left[\left( ( \alpha, \beta) \cdot (z_0, z_1), \left(z_1/|z_1|\right)^{-n} w\right)\right]\right) \nonumber \\
&\stackrel{(\ref{eq:3.7}),(\ref{eq:3.5})}{=}&
 \left([z_0^{'}: z_1^{'}], \left(\frac{\beta z + \bar\alpha}{|\beta z + \bar\alpha|}\right)^n w\right),
\end{eqnarray}
where $z_0^{'} = \alpha z_0 - \bar{\beta}z_1$ and $z_1^{'} = \beta z_0 + \bar{\alpha}z_1$.
Let $s:S^2\rightarrow \mathcal{L}_n$ be a  section of the bundle $\mathcal{L}_n$. Using the local trivializations, we get
local sections ($\gamma=N,S$):
\begin{eqnarray}
s_\gamma:U_\gamma & \longrightarrow & U_\gamma\times \mathbb{C}\nonumber\\
     x  &   \longmapsto   & \varphi_\gamma\comp s.
\end{eqnarray}
These are necessarily of the form $s_\gamma(x)=(x,|\psi_\gamma(x)\rangle)$, with
$x\mapsto |\psi_\gamma(x)\rangle$ a complex-valued function defined on $U_\gamma$.
The local version of (\ref{repre:sec}) is, for $\gamma=N$ and $g=(\alpha,\beta)$:
\begin{eqnarray}
(U_{\tiny{\mbox{loc}}}(g) s_N)(x) &=& \sigma_g\left( s_N(g^{-1}\cdot x)\right) =
\sigma_g\left((g^{-1}\cdot x, \left|\psi_N(g^{-1}\cdot x)\right\rangle)\right) \nonumber \\
&=& \left(x, \left(\frac{\beta z(x) + \bar{\alpha}}{|\beta z(x) + \bar{\alpha}|}\right)^n \left|\psi_N(g^{-1}\cdot
x)\right\rangle\right)\\
&=:&\left(x, \omega(x, g) \left|\psi_N(g^{-1}\cdot
x)\right\rangle\right),\nonumber
\end{eqnarray}
with $\omega(x,g)$ defined through the last equality.
In order to find local expressions for the infinitesimal generators,
 we introduce, for each generator, an appropriate parametrization $t\mapsto g(t)$.
The corresponding generators are then defined   by their action on local sections ($s_N\mapsto J s_N$)
in the following way:
\begin{equation}
(J s_N)(x):=i\frac{d}{dt}\Big|_{t=0} (U_{\tiny{\mbox{loc}}}(g(t)) s_N)(x).
\end{equation}
The general form of the generator will be $J = \tilde{L} + \tilde{\omega}$, where
$\tilde{\omega}$ is a $x$ dependent factor and $\tilde{L}$ a differential operator. This can be seen from
\begin{eqnarray}
\lefteqn{i\frac{d}{dt}\Big|_{t=0} \left(\omega(x,g(t)) \left|\psi_N(g(t)^{-1}\cdot x)\right\rangle\right)= {}}
 \nonumber \\
& &{}= \underbrace{\left(i\frac{d}{dt}\Big|_{t=0} \omega(x,g(t))\right)}_{=\tilde\omega(x)}
 \left|\psi_N(x)\right\rangle +
 \underbrace{i\frac{d}{dt}\Big|_{t=0} \left|\psi_N(g(t)^{-1}\cdot x)\right\rangle}_{=
 \tilde L \left|\psi_N(x)\right\rangle}.
\end{eqnarray}
The generator for rotations around the $z$ axis is obtained by putting $g(t)=(\alpha(t),\beta(t))=(e^{it/2},0)$.
In this case. $\tilde L= \hat L_z$, the third component of the usual  (orbital) angular momentum
operator. For $\tilde \omega$ we obtain
\[
\tilde{\omega}_z(x) = i\frac{d\omega}{dt}\Big|_{t=0} = \frac{n}{2} .
\]
Thus,
\[
\hat{J}^N_z = \hat{L}_z + \frac{n}{2} .
\]
For rotation around the $y$ axis we put $\alpha(t)=\cos t/2$ and $\beta(t)= \sin t/2$. This leads to:
\begin{eqnarray}
\tilde{\omega}_y(x) &=& i\frac{d\omega(x,g(t))}{dt}\Big|_{t=0} =
i\frac{d}{dt} \Big|_{t=0} \left(\frac{\cos t/2 + \sin t/2\, z(x)}{\cos t/2 + \sin t/2\, \overline{z(x)}}\right)^{n/2}
 \nonumber \\
&=& - \frac{n}{4} \left(z(x) - \overline{z(x)}\right) = \frac{n}{2}\frac{\sin\theta\sin\varphi}{1 - \cos\theta} .
\end{eqnarray}
Here, again we have $\tilde L= \hat L_y$, with $\hat L_y$ the second component of the usual  (orbital) angular momentum
operator. It follows that
\begin{equation}
J^N_y = L_y - \frac{n}{2} \frac{y}{1 - z}.
\end{equation}
Using the commutation relations, we obtain, for the remaining generator,
\begin{equation}
J^N_x = L_x - \frac{n}{2} \frac{x}{1 - z}.
\end{equation}
Writing $J^N=(J^N_x,J^N_y,J^N_z)$, we can express the result of the previous computations as follows:
\begin{equation}
J^N = L - \frac{n}{2} K^N.
\end{equation}
Here, $L$ represents the usual orbital  angular momentum operator and $K^N$ is given by (\ref{eq:3.8}) (here it is regarded as
a multiplication operator). The result for the local operator $J^S$ is obtained in the same way.

Comparing with the classical expression (\ref{eq:3.3}) we see that the condition
$\mu = n/2$
must be imposed in order to obtain a consistent quantum theory.
This is, in fact, an expression of the quantization of the electric charge,
obtained by the canonical group quantization method. Notice that here we are only considering
the \emph{kinematical} part of the problem.

\section{Rotation generators for indistinguishable particles}
\label{sec:4}
\subsection{Configuration space}
\label{sec:indistingui}
The configuration space for a system of $N$ indistinguishable, non-colliding particles in $\mathbb{R}^3$
is defined as
\begin{equation}
Q_{N} = \tilde{Q}_{N}/S_{N},
\end{equation}
where
\begin{equation}
\tilde{Q}_{N} = \left\{(r_{1},...,r_{N})\in\Bbb{R}^{3N}\,|\,r_{i}\neq
r_{j} \,  \mbox{ whenever }\, i\neq j\right\},
\end{equation}
with $S_{N}$ denoting the permutation group. We are interested in the case $N=2$, for which
we have the following homeomorphism:
 \begin{equation}
 Q_2 \cong \Bbb{R}^3 \times \Bbb{R}_{+} \times \Bbb{R}P^2.
 \end{equation}
Here, the projective space  $\Bbb{R}P^2$ is obtained, through identification of exchanged configurations,
from the sphere consisting  of all  normalized relative position vectors. Since we are only interested on
 topological effects, we regard $\Bbb{R}P^2$ as the configuration space for this problem. It is well known
 that the quotient map $S^2\rightarrow \Bbb{R}P^2$ gives place to a $\mathbb{Z}_2$-bundle structure and also that
 there are two inequivalent  (scalar)  quantizations  on $\Bbb{R}P^2$, determined by the characters of the
 fundamental group  $\pi_1(\Bbb{R}P^2)\cong \mathbb{Z}_2$. Since our aim is to construct the infinitesimal generators of
 rotations for this problem, it will  convenient to describe the configuration space both as the quotient
 $S^2/\mathbb{Z}_2$  and as  a homogeneous space, of the form $SU(2)/H$. Setting
\begin{equation}
H: = \left\{
\left( \begin{array}{ccc}
\lambda & 0 \\
0 & \bar{\lambda}
\end{array} \right),
\left( \begin{array}{ccc}
0 & \bar{\lambda} \\
-\lambda & 0
\end{array} \right) | \, \, \, |\lambda|^2 = 1  \right\},
\end{equation}
one can show that the space of right orbits of $H$ on $SU(2)$ is homeomorphic to  $\Bbb{R}P^2$.
We will denote the orbits of this action as $[[z_0 : z_1 ]]$, where $[z_0:z_1]\in \mathbb{C}P^1\cong S^2$.
\subsection{Construction of the angular momentum operators}
The construction is similar to the one presented in the previous section. Since the configuration space is of the form
$SU(2)/H$, we start by considering unitary representations of the group $H$. In one complex dimension,
 we only have two possibilities, given by the trivial representation (boson statistics) and by
\begin{eqnarray}
\kappa:\hspace{1cm} H \hspace{0.7cm}&\longrightarrow& \mbox{Gl}(\Bbb{C}) \nonumber \\
\left( \begin{array}{ccc}
\lambda & 0 \\
0 & \bar{\lambda}
\end{array} \right) &\longmapsto& \;\;\;1,  \\
\left( \begin{array}{ccc}
0 & \bar{\lambda} \\
-\lambda & 0
\end{array} \right) &\longmapsto& -1 \nonumber.
\end{eqnarray}
From now on, we will only consider this representation, which is the one giving place to Fermi statistics
 (i.e. wave functions for scalar particles violating the spin-statistics connection).
The total space of the   line bundle $SU(2) \times_{\kappa} \Bbb{C}$ associated to the principal bundle
$SU(2)\rightarrow SU(2)/H$
is the space $\left\{[(g,v)]\, | \, g\in SU(2) \, {\rm and} \, v \in \Bbb{C}\right\}$ of equivalence
classes defined by the equivalence relation $(g,v)\sim (g h,\kappa(h^{-1})v)$. The projection
is given by
\begin{equation}
\pi_\kappa\left([(z_0,z_1),v]\right)=[[z_0:z_1]].
\end{equation}
 The action of the
rotation group $SU(2)$ on the configuration space in the one naturally induced by the action on the sphere. That is, for
$g=(\alpha,\beta)\in SU(2)$ and $p=[[z_0:z_1]]\in SU(2)/H$, we have:
 \begin{equation}
l_g(p)=[[ \alpha z_0 - \bar{\beta}z_1: \beta z_0 + \bar{\alpha} z_1]].
\end{equation}
As in the previous section, the action can be lifted to the total space of the bundle, by setting
\begin{equation}
l^\uparrow_g\left([(z_0,z_1),v]\right)= \left[\left(g(z_0,z_1), v\right)\right].
\end{equation}

In the example of the magnetic monopole
we had to introduce local trivializations in order to obtain the known expressions for the  angular momentum
operators. For  the case of indistinguishable particles that we are considering in this section,
our purpose is to establish a bridge between our formalism and the one presented in
\cite{Berry1997}. The latter does not  make explicit use of vector bundles. Instead, it uses a
 position dependent spin basis. The spin basis vectors
are actually sections of a trivial bundle on the sphere, but their transformation properties allow one to regard
wave functions constructed from them as sections on a bundle over the physical configuration
 space\footnote{There are certain subtleties involved in this identification,
  that have been discussed in \cite{Reyes:06}.}. So, in order to establish this connection between the two formalisms,
  we will construct an explicit isomorphism between the bundle
$SU(2)\times_\kappa \mathbb{C}$ and a line subbundle $\mathcal{L}_-$ of the  trivial bundle
$\mathbb{R}P^2\times \mathbb{C}^3\rightarrow \mathbb{R}P^2$, as described below.

Let us regard the projective plane as the quotient $S^2/\mathbb{Z}_2$. Then, points on it are equivalence
classes of the form $[x]=\lbrace x,-x\rbrace$, where $x=(x_1,x_2,x_3)\in S^2$. With this,   the  following open cover
 can be defined ($\alpha=1,2,3$):
\begin{equation}
\label{cartasloc}
U_{\alpha} =  \{[x] \in \mathbb{R}P^2\, |\; x_\alpha \neq 0\}.
\end{equation}

Let us now  define a line  bundle $\mathcal{L}_{-}$ (a sub-bundle of the
trivial bundle $\mathbb{R}P^2 \times \Bbb{C}^3$) as follows. The total space of the bundle is given by the following
set:
\begin{equation}
\label{eq:4.1}
\left\{\left(\left[x\right], \lambda\left|\phi(x)\right\rangle\right)\in \mathbb{R}P^2 \times \Bbb{C}^3 \, | \, \lambda \in \Bbb{C} \, {\rm and} \, x \in \left[x\right]\right\},
\end{equation}
where
\begin{eqnarray}
 |\phi(-)\rangle : S^2 & \longrightarrow & \mathbb{C}^3 \nonumber\\
 x &\longmapsto &|\phi(x)\rangle
\end{eqnarray}
is \emph{any} map from $S^2$ to  $\mathbb{C}^3$ satisfying the following conditions:
\begin{itemize}
\item[({\it i})] It is smooth.
\item[({\it ii})] $|\phi(x)\rangle\neq  0$ for all $x\in S^2$.
\item[({\it iii})] $|\phi(-x)\rangle = -|\phi(x)\rangle$ for all $x\in S^2$.
\end{itemize}
The bundle projection is defined through
 $\pi\left(\left(\,\left[x\right], \lambda\left|\phi(x)\right\rangle\,\right)\right)=\left[x\right]$.
According to (\ref{eq:4.1}), an element in the total space of $\mathcal{L}_-$ is given by a tuple
of the form $([x],\lambda|\phi(x)\rangle)$. Notice that there is some ambiguity in this expression,
since a representative $x$ is being explicitly used. However, there is no problem if one realizes that a choice
of representative $x\in[x]$ uniquely fixes the value of $\lambda$. Assuming that the representative $x$ has been chosen,
and that to it corresponds the scalar $\lambda$, then, from property ({\it iii}) above, it follows
that the other choice of representative, $-x$, forces the value of the scalar to be $-\lambda$. An alternative
 way to define the bundle is by saying that the fiber over the point $[x]$ is the subset
  $\lbrace[x]\rbrace\times V_{[x]}$ of $\mathbb{R}P^2\times \mathbb{C}^3$, where $V_{[x]}$ is the vector
  space generated by the vector $|\phi(x)\rangle\in \mathbb{C}^3$.
Local trivializations for $\mathcal{L}_-$ are given by ($\alpha=1,2,3$):
\begin{eqnarray}
\varphi_\alpha : \pi^{-1}(U_\alpha) &\longrightarrow& U_\alpha \times \Bbb{C} \nonumber \\
\left(\left[x\right], \lambda\left|\phi(x)\right\rangle\right) &\longmapsto& \left(\left[x\right], {\rm sign}(x_\alpha)\lambda\right).
\end{eqnarray}
They give place to the following transition functions:
\begin{eqnarray}
g_{\alpha\beta}: U_\alpha \cap U_\alpha & \longrightarrow & \mathbb{Z}_2\leqslant U(1)\nonumber\\
\left[\,x\right]&\longmapsto &g_{\alpha\beta}([x])= \mbox{sign}(x_\alpha x_\beta).
\end{eqnarray}
  Yet another point of view
  is provided by the Serre-Swan equivalence of bundles and modules: given a (normalized) map $|\phi(-)\rangle$ satisfying
  properties ({\it i})-({\it iii}), it can be shown that the projector $p:[x]\mapsto |\phi(x)\rangle\langle\phi(x)|$
gives place to a finitely generated projective module $p(\mathcal{A}_+^3)$ over the algebra $\mathcal{A}_+$ of complex, continuous
even functions over the sphere \cite{Paschke:01,Papadopoulos:04}. This module is isomorphic to
 the module  of sections on the bundle
$\mathcal{L}_-$.

If  $g=(z_0,z_1)\in SU(2)$ and $v\in\mathbb{C}$,
then $\pi_\kappa([(g,v)])=[[z_0:z_1]]$ is a point in $SU(2)/H$. Let $x(g)$ denote the point in $S^2$ obtained
from $g$ through the quotient map $SU(2)\rightarrow SU(2)/U(1)$ and let $[x(g)]$ denote the corresponding
equivalence class, with respect to the quotient map $S^2\rightarrow S^2/\mathbb{Z}_2$. Then it is clear
that $\pi_\kappa([(g,v)])=[x(g)]$, independently of the  chosen $g$. This fact allows us to construct
the following map between the total spaces of $SU(2)\times_\kappa \mathbb{C}$ and $\mathcal{L}_-$:
\begin{eqnarray}
\Phi:   SU(2) \times_{\kappa} \Bbb{C} &\longrightarrow& \;\;\;\;\;\mathcal{L}_{-} \nonumber\\
\left[(g, v)\right]\;\;\; &\longmapsto& \left(\left[x(g)\right], v\left|\phi(x(g))\right\rangle\right).
\end{eqnarray}
It is easy to check that this map is well defined and that, in fact, it provides a bundle isomorphism. Thus, we
obtain an induced lift on the bundle $\mathcal{L}_-$, as indicated in the following diagram:
\begin{equation}
\xymatrix{
  & & & \mathcal{L}_{-} \ar @{-->}[rr]^{\tau_g}
 \ar[d]_{\Phi^{-1}}
 &
  & \mathcal{L}_{-}  & &\\
  & & & {SU(2) \times_{\mathcal{U}_n} \Bbb{C}}\ar[rr]^{l^{\uparrow}_g}
 \ar[d]^{\pi_\kappa}
 &
  & SU(2) \times_{\mathcal{U}_n} \Bbb{C} \ar[u]^{\Phi} \ar[d]_{\pi_\kappa} & & &\\
   & & &
\mathbb{R}P^2
\ar[rr]^{l_g}& &
\mathbb{R}P^2  & & \\
    }
\end{equation}
From $\tau_g = \Phi \circ l^{\uparrow}_g \circ \Phi^{-1}$ we get, for $g=(\alpha,\beta)$:
\begin{eqnarray}
\tau_g\left(\left[x\right], \lambda \left|\phi(x)\right\rangle\right) &=&
(\Phi \circ l^{\uparrow}_g \circ \Phi^{-1}) \left(\left[x\right], \lambda \left|\phi(x)\right\rangle\right) \nonumber \\
&=& (\Phi \circ l^{\uparrow}_g ) \left[(\left(z_0, z_1\right), \lambda)\right] \nonumber \\
&=& \Phi \left[(\left(\alpha, \beta\right) \cdot \left(z_0, z_1\right), \lambda)\right] \nonumber \\
&=& \left(\left[x(z^{'}_0, z^{'}_1)\right], \lambda \left|\phi(x(z^{'}_0, z^{'}_1))\right\rangle\right),
\end{eqnarray}
where, as in the previous section, $z^{'}_0 = \alpha z_0 - \bar{\beta}z_1$ and
$z^{'}_1 = \beta z_0 + \bar{\alpha} z_1$. Here, $(z_0,z_1)$ is chosen in such a way that $x\equiv x(z_0,z_1)=[[z_0:z_1]]$.

Now, notice that a smooth section on $\mathcal{L}_-$ can always be written
in the form $\Psi([x]) = (\left[x\right], a(x) \left|\phi(x)\right\rangle)$,
with $a:S^2\rightarrow \mathbb{C}$ a smooth \emph{antisymmetric} function.
Such a section transforms under the action of
$SU(2)$ in the following way:
\begin{eqnarray}
(U(g)\Psi)([x]) &:=& \tau_g(\Psi(g^{-1}\cdot [x])) = \tau_g(\left[g^{-1}\cdot x\right],
 a(g^{-1}\cdot x) \left|\phi(g^{-1}\cdot x)\right\rangle) \nonumber \\
&=& \left(\left[x\right], a(g^{-1}\cdot x) \left|\phi(x)\right\rangle\right).
\end{eqnarray}
From this we immediately see that the infinitesimal generators $J_i$ are given by
\begin{equation}
\label{eq:4.2}
(J_i\Psi)([x]) = \left(\left[x\right], (L_i a)( x) \left|\phi(x)\right\rangle\right),
\end{equation}
where $L_i$ is the usual (orbital) angular momentum operator.

\section{Conclusions}
The generally accepted (relativistic) quantum field theory proof of the Spin-Statistics Theorem is
perhaps one of the most interesting results of the general theory of quantum fields and there should be
no apparent reason for trying to look for a different proof. But,
as a brief look at the current literature on the subject will show, the interest in the problem of
the spin-statistics connection in non relativistic quantum mechanics
has increased in the last years. One reason might be that there is the opinion that
non relativistic quantum mechanics describes, without  relativity, an astonishing amount of physical phenomena.
Being a theory that stands on a firm mathematical foundation, one would like to be able to obtain
the physically correct spin-statistics connection without having to draw a theorem from another theory
(which, anyway, is more fundamental). Another motivation would be the study of the spin-statistics connection
in different contexts: quantum gravity, quantum field theory on non-commutative spaces, etc..
In any case, in contrast to the opinion of many authors, our interest is not so much
to find a \emph{simple proof} of the theorem, or even one which does not use relativistic invariance, but
rather to understand the connection from a different point of view. Just the fact that the Fermi-Bose alternative
can be obtained as a consequence of the topology of the configuration space is a quite remarkable result. But,
if in the end
it turns out that the connection has something to do with topology or geometry, one should not expect to obtain
an understanding of it without using the tools of geometry and topology. The approach that we are presenting
here, which in some aspects is a continuation of \cite{Papadopoulos:04}, has the purpose
of establishing a bridge between the proposed mathematical/physical framework and the current literature
on the subject. We believe  that a clear formulation of the problem in mathematical terms might help in providing
a firm foundation to many works where interesting physical ideas have been put forward and to establish a link between them.

The main result of the present paper is the construction of the angular momentum operators for a system of two
indistinguishable particles obeying fermionic statistics. Taking into account the equivalence $\Gamma(\mathcal{L}_-)\cong
\mathcal{A}_-$ \cite{Paschke:01},\cite{Reyes:06} we see from (\ref{eq:4.2}) that, not only sections on $\mathcal{L}_-$
can be isomorphically mapped to antisymmetric functions over $S^2$, but also that the generators of rotations, obtained here
by means of   a well defined quantization map, correspond to the \emph{usual} angular momentum operators. Thus, whereas
it is true that by taking seriously
into account the indistinguishability of quantum particles
we are forced to consider non trivial geometric/topological structures, at the end we see that all these structures
can be mapped isomorphically to the ones that we are ``familiar'' with. One could argue that this only means we have not
won anything. On the contrary, we believe that taking these structures into account could eventually lead to an advance
in our understanding of the subject. In particular, we believe that it would be a fruitful idea
to obtain a global version of the theorem proven in \cite{Kuckert:04}, using the tools discussed in the present paper.

\bibliographystyle{amsalpha}

\begin{thebibliography}{ZSN{\etalchar{+}}83}

\bibitem[AB02]{Atiyah2002a}
M.~Atiyah and R.~Bielawski, \emph{Nahm's equations, configurations spaces and
  flag manifolds}, Bull. Brazilian Math. Soc. \textbf{33} (2002), 157--166.

\bibitem[AM03]{Allen:03}
R.E. Allen and A.R. Mondragon, \emph{Comment on ``spin and statistics in
  nonrelativistic quantum mechanics: The spin-zero case"}, Phys. Rev. A
  \textbf{68} (2003), 046101.

\bibitem[Ana02]{Anastopoulos:02}
Charis Anastopoulos, \emph{Spin-statistics theorem and geometric quantization},
  Int. J. Mod. Phys. A \textbf{19} (2002), 655--676.

\bibitem[AS02]{Atiyah2002}
M.~Atiyah and P.~Sutcliffe, \emph{The geometry of point particles}, Proc. R.
  Soc. London A \textbf{458} (2002), 1089--1115.

\bibitem[Ati01]{Atiyah2001}
M.~Atiyah, \emph{Configurations of points}, Phil. Trans. Roy. Soc. Lond. A359
  \textbf{359} (2001), no.~1784, 1375--1387.

\bibitem[BL81]{Biedenharn1981}
L.C. Biedenharn and J.D. Louck, \emph{The racah-wigner algebra in quantum
  theory}, Encyclopedia of Mathematics and its Applications-Vol. 9 (G.C. Rota,
  ed.), Addison-Wesley, 1981.

\bibitem[BR97]{Berry1997}
M.V. Berry and J.M. Robbins, \emph{Indistinguishability for quantum particles:
  spin, statistics and the geometric phase}, Proc. R. Soc. London A
  \textbf{453} (1997), 1771--1790.

\bibitem[BR00]{Berry:00}
\bysame, \emph{Quantum indistinguishability: alternative constructions of the
  transported basis}, J. Phys. A \textbf{33} (2000), L207--L214.

\bibitem[CJ04]{Chruscinski:04}
Dariusz Chru{\'s}ci{\'n}ski and Andrzej Jamio{\l}kowski, \emph{Geometric phases
  in classical and quantum mechanics}, Birkhäuser, Boston, 2004.

\bibitem[Dir31]{Dirac:31}
P.A.M. Dirac, \emph{Local observables and particle statistics ii}, Proc. Roy.
  Soc. \textbf{A 133} (1931), 60.

\bibitem[DS98a]{Duck:97}
I.~Duck and E.C.G. Sudarshan, \emph{Toward an understanding of the
  spin-statistics theorem}, Am. J. Phys. \textbf{66} (1998), 284--303.

\bibitem[DS98b]{Duck1998}
Ian Duck and E.C.G. Sudarshan, \emph{Pauli and the spin-statistics theorem},
  World Scientific, 1998.

\bibitem[Fie39]{Fierz:39}
M.~Fierz, \emph{On the relativistic theory of force free particles of arbitrary
  spin}, Helv. Phys. Acta \textbf{12} (1939), 3--37.

\bibitem[FR68]{Finkelstein:68}
David Finkelstein and Julio Rubinstein, \emph{Connection between spin,
  statistics, and kinks}, J. Math. Phys. \textbf{9} (1968), 1762--1779.

\bibitem[Haa96]{Haag1996}
Rudolf Haag, \emph{Local quantum physics}, 2nd ed., Texts and Monographs in
  Physics, Springer Verlag, 1996.

\bibitem[Ish84]{Isham:84}
C.~J. Isham, \emph{Topological and global aspects of quantum theory},
  Relativity, Groups and Topology II (Amsterdam) (Bryce~S. DeWitt and Raymond
  Stora, eds.), Kluwer academic publishers, 1984, pp.~1059--1290.

\bibitem[Kas06]{Kastrup2006a}
H.A. Kastrup, \emph{Quantization of the canonically conjugate pair angle and
  orbital angular momentum}, Physical Review A \textbf{73} (2006), 052104.

\bibitem[Kir76]{Kirillov1976}
A.A. Kirillov, \emph{Elements of the theory of representations}, Springer
  Verlag, 1976.

\bibitem[Kuc04]{Kuckert:04}
B.~Kuckert, \emph{Spin and statistics in nonrelativistic quantum mechanics, i},
  Physics Letters A \textbf{322} (2004), 47--53.

\bibitem[LD71]{Laidlaw:71}
Michael~G.G. Laidlaw and Cecile~Morette DeWitt, \emph{Feynman functional
  integrals for systems of indistinguishable particles}, Phys. Rev. D
  \textbf{3} (1971), 1375--1378.

\bibitem[LM77]{Leinaas:77}
J.M. Leinaas and J.~Myrheim, \emph{On the theory of identical particles}, Il
  Nuovo Cimento \textbf{37B} (1977), 1--23.

\bibitem[Mac68]{Mackey1968}
G.~W. Mackey, \emph{Induced representations and quantum mechanics}, W. A.
  Benjamin, New York, 1968.

\bibitem[Pas01]{Paschke:01}
M.~Paschke, \emph{Von nichtkommutativen geometrien, ihren symmetrien und etwas
  hochenergiephysik}, Ph.D. thesis, Institut f\"ur Physik, Universit\"at Mainz,
  2001.

\bibitem[Pau40]{Pauli:40}
W.~Pauli, \emph{The connection between spin and statistics}, Phys. Rev.
  \textbf{58} (1940), 716--722.

\bibitem[Pes03a]{Peshkin2003a}
Murray Peshkin, \emph{Reply to comment on spin and statistics in
  nonrelativistic quantum mechanics: The spin-zero case}, Phys. Rev. A
  \textbf{68} (2003), 046102.

\bibitem[Pes03b]{Peshkin2003}
\bysame, \emph{Spin and statistic in nonrelativistic quantum mechanics: The
  spin-zero case}, Phys. Rev. A \textbf{67} (2003), 042102.

\bibitem[PPRS04]{Papadopoulos:04}
N.A. Papadopoulos, M.~Paschke, A.~Reyes, and F.~Scheck, \emph{The
  spin-statistics relation in nonrelativistic quantum mechanics and projective
  modules}, Annales Math\'ematiques Blaise-Pascal \textbf{11} (2004), no.~2,
  205--220.

\bibitem[PRL]{Papadopoulos}
N.A. Papadopoulos and A.F. Reyes-Lega, \emph{Quantum indistinguishability and
  single-valuedness: a new approach to an old problem}, In preparation.

\bibitem[Rey06]{Reyes:06}
A.~Reyes, \emph{On the geometry of the spin-statistics connection in quantum
  mechanics}, PhD Dissertation, Mainz, 2006.

\bibitem[Sch68]{Schulman:68}
Lawrence Schulman, \emph{A path integral for spin}, Phys. Rev. \textbf{5}
  (1968), 1558--1569.

\bibitem[SD03]{Sudarshan:03}
E.C.G. Sudarshan and I.M. Duck, \emph{What price the spin-statistics theorem?},
  Pramana Journal of Physics \textbf{61} (2003), 1--9.

\bibitem[Sud75]{Sudarshan1975}
E.C.G. Sudarshan, \emph{Relation between spin and statistics}, Stat. Phys.
  Suppl.: J. Indian Inst. Sci. \textbf{June} (1975), 123--137.

\bibitem[SW00]{WightmanCPT:00}
R.F. Streater and A.S. Wightman, \emph{Spin, statistics and all that},
  Princeton U. Press, Princeton, 2000.

\bibitem[Wil82]{Wilczek1982}
Frank Wilczek, \emph{Quantum mechanics of fractional-spin particles}, Phys.
  Rev. Lett. \textbf{49} (1982), 957 -- 959.

\bibitem[Woo80]{woodhouse:80}
N.~Woodhouse, \emph{Geometric quantization}, Clarendon Press, Oxford, 1980.

\bibitem[ZSN{\etalchar{+}}83]{Zaccaria1983}
F.~Zaccaria, E.C.G. Sudarshan, J.S. Nilsson, N.~Mukunda, G.~Marmo, and A.P.
  Balachandran, \emph{Universal unfolding of hamiltonian systems: From
  symplectic structure to fiber bundles}, Physical Review D \textbf{27} (1983),
  2327--2340.

\end{thebibliography}
\newcommand{\etalchar}[1]{$^{#1}$}
\providecommand{\bysame}{\leavevmode\hbox to3em{\hrulefill}\thinspace}
\providecommand{\MR}{\relax\ifhmode\unskip\space\fi MR }
\providecommand{\MRhref}[2]{%
  \href{http://www.ams.org/mathscinet-getitem?mr=#1}{#2}
} \providecommand{\href}[2]{#2}

\end{document}